\documentclass[a4paper,11pt]{article}
\usepackage{jheppub} % for details on the use of the package, please see the JINST-author-manual
\usepackage{lineno}
\usepackage{physics}
\usepackage{amsmath}
\usepackage{amssymb}
\usepackage{graphicx}
\usepackage{overpic}
\usepackage{authblk}
\usepackage{epsfig}
\usepackage{epstopdf}
\usepackage{colortbl}
\usepackage{rotating}
\usepackage{bm} % bold math
\usepackage{dcolumn}% Align table columns on decimal point
\usepackage{multirow}
\usepackage{makecell}
\usepackage{indentfirst}
\usepackage{booktabs}
\usepackage{arydshln}
\usepackage{xcolor}

\newcommand{\BESIIIorcid}[1]{\href{https://orcid.org/#1}{\hspace*{0.1em}\raisebox{-0.45ex}{\includegraphics[width=1em]{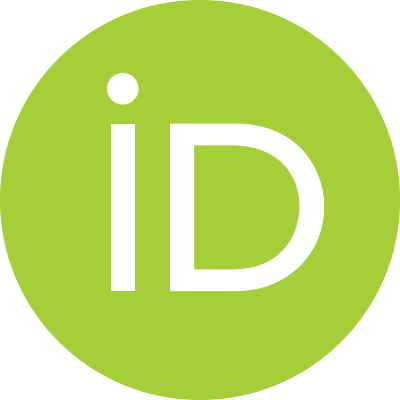}}}} 

\title{\boldmath Search for the charmonium weak decay $J/\psi\to\bar{D}^0\bar{K}^{*0}+{\rm c.c.}$}

\collaboration{
BESIII Collaboration\hyperref[author]{*}\\
\includegraphics[height=15mm]{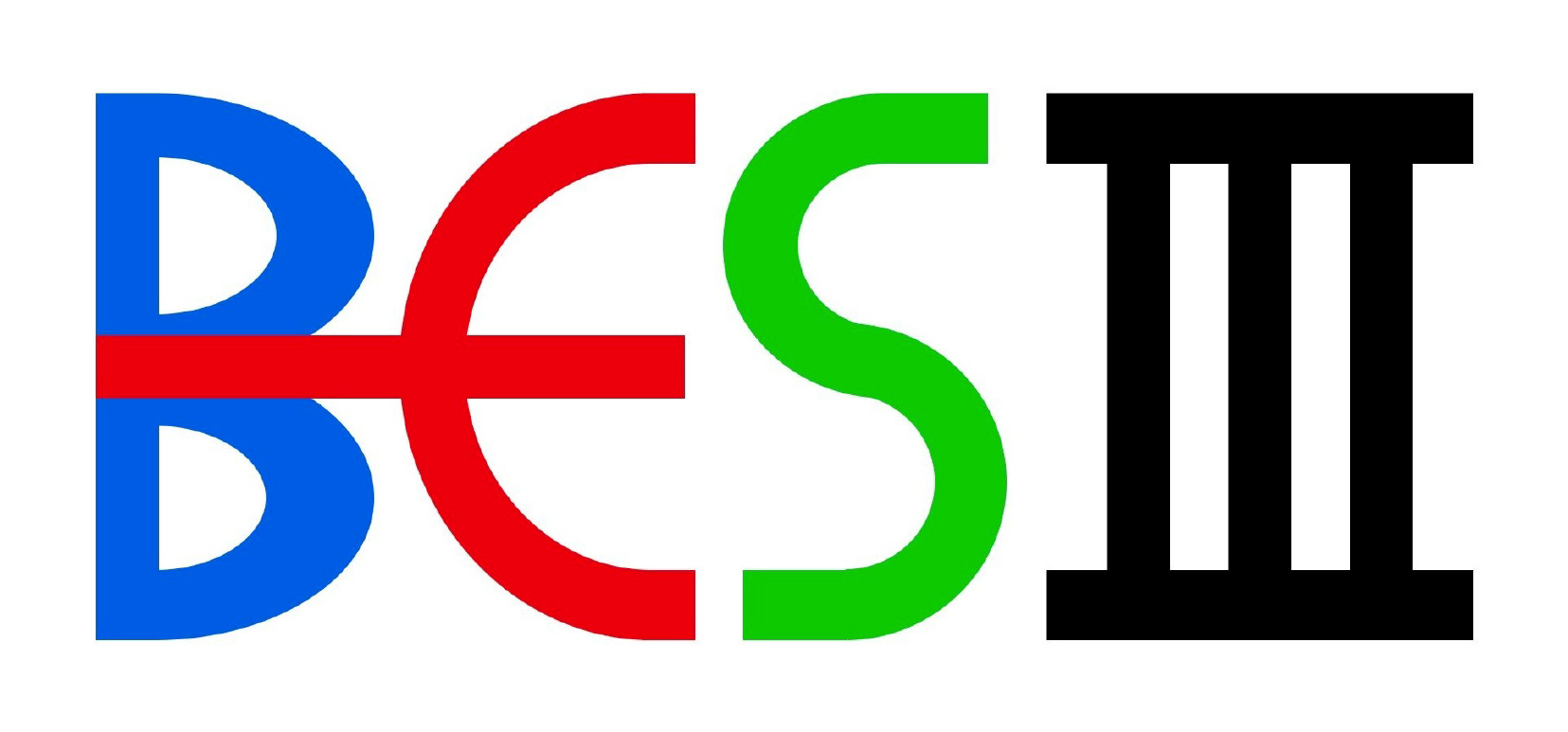} 
}

\emailAdd{besiii-publications@ihep.ac.cn}

\abstract{
Based on a sample of $(10087\pm44)\times10^6$ $J/\psi$ events collected at the center-of-mass energy $\sqrt{s}$ = 3.0969 GeV with the BESIII detector, we search for the charmonium rare weak decay $J/\psi\to\bar{D}^0\bar{K}^{*0}+{\rm c.c.}$. No significant signal is observed, and the upper limit on its decay branching fraction at the 90\% confidence level is set as $1.4\times10^{-7}$, improving the sensitivity of the previous best limit by an order of magnitude.
}

\keywords{$e^+e^-$ experiments, charmonium, weak decay}

\begin{document}
\maketitle
\flushbottom

%%====================================================

\section{Introduction}
\label{sec:intro}

The charmonium state $J/\psi$, which primarily decays via strong and electromagnetic interactions, has been extensively studied for decades. Since the $J/\psi$ mass lies below the open charm threshold, it cannot decay into a pair of charmed mesons. However, weak decays of $J/\psi \to D_{(s)}^{(*)}X$ are kinematically allowed, where $X$ denotes light hadrons or light lepton pairs such as $e\bar{\nu}_e$ or $\mu\bar{\nu}_\mu$. Throughout this paper, charge-conjugate processes are implied.
	
\begin{figure}[!ht]
    \centering
    \includegraphics[width=0.4\textwidth]{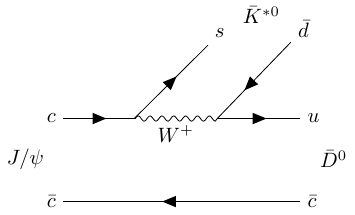}
    \caption{The Feynman diagram of $J/\psi\to\bar{D}^0\bar{K}^{*0}$ decay.}
    \label{fg:1.1}
\end{figure}

Within the Standard Model (SM), the inclusive branching fraction (BF) of the $J/\psi$ rare weak decays containing one charmed meson is predicted to be about $10^{-8}$ or lower~\cite{bib:1-1, bib:1-2, bib:1-3, bib:1-4, bib:1-5, bib:1-6, bib:1-7, bib:1-8, bib:1-9, bib:1-10, bib:1-10.1}. There have been some theoretical predictions for various exclusive decay modes, including the hadronic decays $J/\psi \to PP, PV$ (where $P$ represents a pseudoscalar meson and $V$ a vector meson)~\cite{bib:1-8,bib:1-5,bib:1-7,bib:1-4}, and some semi-leptonic decays~\cite{bib:1-6,bib:1-7,bib:1-8,bib:1-9,bib:1-10,bib:1-25.6}. However, none of these decays have yet been observed~\cite{bib:1-11,bib:1-12,bib:1-13,bib:1-14,bib:1-15,bib:1-10.2,bib:1-15.1}. A summary of the experimental results and the SM predictions for $J/\psi$ rare weak decays is presented in Table~\ref{tb:1}. Figure~\ref{fg:1.1} shows the tree-level Feynman diagram for the weak decay $J/\psi\to\bar{D}^0\bar{K}^{*0}$.

\begin{table}[!ht]
    \centering
    \caption{Experimental results and SM predictions for the BFs of $J/\psi$ weak decays. For each decay mode, the total size of the $J/\psi$ sample used, the upper limit (UL) on the BF at 90\% confidence level (CL), and the BF predicted by the SM $(\mathcal{B}_{\rm SM})$ are given.}
    \label{tb:1}
    \begin{tabular}{clccc}
    \toprule
    Decay type & \multicolumn{1}{c}{Decay mode} & $N_{J/\psi}$ $(\times10^6)$ & $\rm UL$ at 90\% CL & $\mathcal{B}_{\rm SM}$ $(\times10^{-10})$\\
    \midrule
    $c\to s$ &&&& \cite{bib:1-8,bib:1-5,bib:1-7,bib:1-4} \\
    $PP$ & $J/\psi\to D_s^-\pi^+$ & 58 & $1.4\times10^{-4}$~\cite{bib:1-11} & 2.00$\sim$8.74 \\
    &$J/\psi\to \bar{D}^0\bar{K}^0$ & 58 & $1.7\times10^{-4}$~\cite{bib:1-11} & 0.36$\sim$2.80 \\
    $PV$ &$J/\psi\to D_s^-\rho^+$ & 225.3 & $1.3\times10^{-5}$~\cite{bib:1-12} & 12.60$\sim$50.50 \\
    & $J/\psi\to \bar{D}^0\bar{K}^{*0}$ & 225.3 & $2.5\times10^{-6}$~\cite{bib:1-12} & 1.54$\sim$10.27 \\
    \cmidrule(lr){1-5}
    $c\to d$ &&&& \cite{bib:1-8,bib:1-5,bib:1-7,bib:1-4} \\
    $PP$ & $J/\psi\to D^-\pi^+$ & 10087 & $7.0\times10^{-8}$~\cite{bib:1-10.2} & $0.08\sim0.55$ \\
    &$J/\psi\to D^0\pi^0$ & 10087 & $4.7\times10^{-7}$~\cite{bib:1-10.2} & $0.024\sim0.055$ \\
    &$J/\psi\to D^0\eta$ & 10087 & $6.8\times10^{-7}$~\cite{bib:1-10.2} & $0.016\sim0.070$ \\
    $PV$ &$J/\psi\to D^-\rho^+$ & 10087 & $6.0\times10^{-7}$~\cite{bib:1-10.2} & $0.42\sim2.20$ \\
    &$J/\psi\to D^0\rho^0$& 10087 & $5.2\times10^{-7}$~\cite{bib:1-10.2} & $0.18\sim0.22$ \\
    \midrule
    \multicolumn{2}{l}{Semi-leptonic} &&&~\cite{bib:1-6,bib:1-7,bib:1-8,bib:1-9,bib:1-10,bib:1-25.6}\\
    $c\to s$&$J/\psi\to D_s^-e^+\nu_e$ & 225.3 & $1.3\times10^{-6}$~\cite{bib:1-13} & $1.8\sim10.4$ \\
    &$J/\psi\to D_s^{*-}e^+\nu_e$ & 225.3 & $1.8\times10^{-6}$~\cite{bib:1-13} & $\sim$100 or below \\
    $c\to d$ &$J/\psi\to D^-e^+\nu_e$ & 10087 & $7.1\times10^{-8}$~\cite{bib:1-15} & $0.073\sim0.60$ \\
    &$J/\psi\to D^-\mu^+\nu_\mu$ & 10087 & $9.3\times10^{-7}$~\cite{bib:1-15.1} & $0.071\sim0.58$ \\
    $c\to u$&$J/\psi\to D^0e^+e^-$ & 1310.6 & $8.5\times10^{-8}$~\cite{bib:1-14} & --- \\
    & $J/\psi\to D^0\mu^+\mu^-$ & 10087 & $1.1\times10^{-7}$~\cite{bib:1-14.1} & --- \\
    \bottomrule
    \end{tabular}
\end{table}
	
Searches for these decays serve as tests of the SM, and they also provide a probe of new physics models~\cite{bib:1-16}, such as the top-color model~\cite{bib:1-20}, the minimal supersymmetric standard model with or without $R$-parity violation~\cite{aulakh1982neutrino}, and the two Higgs doublet model~\cite{glashow1977natural}. These models could enhance the BFs of inclusive $J/\psi$ weak decays up to $\mathcal{O}(10^{-5})$~\cite{bib:1-19}.
	 
In this study, we present a search for the weak decay $J/\psi\to\bar{D}^0\bar{K}^{*0}$ utilizing about 10 billion $J/\psi$ events~\cite{bib:3-1} collected with the BESIII detector at the BEPCII collider. A stepwise unblinding analysis strategy is employed to avoid involuntary bias. The selection criteria are determined solely relying on the signal and inclusive Monte Carlo (MC) samples. A randomly selected 10\% of the total data sample is then used to validate the effectiveness of the analysis method and to finalize the procedure of event selection. The same analysis procedure is then applied to the total data sample.  

%%===================================================

\section{BESIII detector and Monte Carlo simulation}
\label{sec:bes}

The BESIII detector~\cite{Ablikim:2009aa} records symmetric $e^+e^-$ collisions provided by the BEPCII storage ring~\cite{Yu:IPAC2016-TUYA01} in the $\tau$-charm energy region. It has collected a large dataset in this energy range~\cite{bib:1-16, Liao:2025lth}. The cylindrical core of the BESIII detector, covering 93\% of the full solid angle, consists of a helium-based multilayer drift chamber (MDC), a plastic scintillator time-of-flight system (TOF), and a CsI(Tl) electromagnetic calorimeter (EMC). All components are enclosed within a $1.0~\rm T$ superconducting solenoidal magnet.  The magnetic field was $0.9~\rm T$ in 2012, corresponding to 11\% of the total $J/\psi$ data. The solenoid is supported by an octagonal flux-return yoke with resistive plate counter muon identification modules interleaved with steel. 
	
The charged-particle momentum resolution at $1~{\rm GeV}/c$ is $0.5\%$, and the energy loss $({\rm d}E/{\rm d}x)$ resolution is 6\% for electrons from Bhabha scattering. The EMC measures photon energies with a resolution of 2.5\% (5\%) at $1~\rm GeV$ in the barrel (end cap) region. The time resolution in the TOF barrel region is $68~\rm ps$, while that in the end cap region was $110~\rm ps$. The end cap TOF system was upgraded in 2015 using multi-gap resistive plate chamber technology, providing a time resolution of $60~\rm ps$, which
benefits 87\% of the data used in this analysis~\cite{etof-1,etof-2,etof-3}.

MC simulated data samples produced with a \textsc{GEANT4}-based~\cite{geant4} software package, which includes the geometric description of the BESIII detector and the detector response, are used to determine detection efficiencies and to estimate backgrounds. The simulation models the beam energy spread and initial state radiation (ISR) in the $e^+e^-$ annihilations with the generator \textsc{KKMC}~\cite{ref:kkmc-1,ref:kkmc-2}. The inclusive MC sample includes both the production of the $J/\psi$ resonance and the continuum processes incorporated in \textsc{KKMC}. All particle decays are modelled with \textsc{EVTGEN}~\cite{ref:evtgen-1,ref:evtgen-2} using branching fractions either taken from the Particle Data Group~\cite{pdg}, when available, or otherwise estimated with \textsc{LUNDCHARM}~\cite{ref:lundcharm-1,ref:lundcharm-2}. Final state radiation~(FSR) from charged final state particles is incorporated using the \textsc{PHOTOS} package~\cite{photos2}.

About 10 billion official $J/\psi$ inclusive MC events corresponding to real data are used to study the backgrounds from $J/\psi$ decays. Additionally, 0.8 million signal MC events are generated to estimate the signal detection efficiency. The process $J/\psi\to\bar{D}^0\bar{K}^{*0}$ is generated with the VVS\_PWAVE model~\cite{ref:evtgen-2}, which simulates $V\to VS$ decays, where $S$ denotes a scalar or pseudo-scalar meson and $V$ denotes a vector meson. The process $\bar{K}^{*0}\to K^-\pi^+$ is generated with the VSS model~\cite{ref:evtgen-2}, which simulates $V\to SS$ decays. The process $\bar{D}^0\to K^+e^-\bar{\nu}_e$ is generated with the SLBKPOLE model~\cite{ref:evtgen-1} , which simulates semi-leptonic decays of $D$ and $B$ mesons. These three models are implemented in EVTGEN~\cite{ref:evtgen-1,ref:evtgen-2}.

%%====================================================

\section{Event selection}
\label{sec:ana}

The analysis is performed using the BESIII offline software system~\cite{bib:3-2}. In the signal process $J/\psi\to \bar{D}^0\bar{K}^{*0}, \bar{D}^0\to K^+e^-\bar{\nu}_e, \bar{K}^{*0}\to K^-\pi^+$, all final-state particles except the neutrino are detected. Since reconstructing the $\bar{D}^0$ meson via non-leptonic decays suffers from poor sensitivity due to significant backgrounds from hadronic $J/\psi$ decays, we opt to reconstruct $\bar{D}^0$ candidates via the semi-leptonic decay $\bar{D}^0\to K^+e^-\bar{\nu}_e$, which has a relatively large BF of $(3.549\pm0.026)\%$~\cite{pdg} and a low background level. 

Charged tracks detected in the MDC are required to fall within the $\theta$ polar angle range such that $\abs{\cos\theta}<0.93$, where $\theta$ is defined with respect to the $z$-axis, the symmetry axis of the MDC. For all charged tracks, the distance of closest approach to the interaction point must be within $10~\rm cm$ along the $z$-axis and within $1~\rm cm$ in the transverse plane. Only events with exactly four selected charged tracks and zero net charge are retained for further analysis.

Particle identification (PID) for charged tracks combines measurements of the specific ionization energy loss in the MDC and the flight time in the TOF to form likelihoods $\mathcal{L}(h)~(h= K,\pi)$ for each hadron $h$ hypothesis. Charged kaon candidates are identified by requiring $\mathcal{L}(K)>0$ and $\mathcal{L}(K)>\mathcal{L}(\pi)$, while charged pion candidates must satisfy $\mathcal{L}(\pi)>0$, $\mathcal{L}(\pi)>\mathcal{L}(K)$ and $\mathcal{L}(\pi)>\mathcal{L}(e)$. Electron identification additionally incorporates information from the EMC. Its likelihoods are required to satisfy $\mathcal{L}(e)>0.001$, $\mathcal{L}(e)/[\mathcal{L}(e)+\mathcal{L}(\pi)+\mathcal{L}(K)] > 0.8$. Electron candidates are further required to have $E/p > 0.86$, where $E/p$ is the ratio of the deposited energy in the EMC to the MDC track momentum.

Photon candidates are identified based on showers in the EMC. The deposited energy of each shower must exceed $25~\rm MeV$ in the barrel region $(\abs{\cos\theta} < 0.80)$ or $50~\rm MeV$ in the end cap region $(0.86 < \abs{\cos\theta}< 0.92)$. To suppress electronic noise and unrelated showers, the difference between the shower time and the event start time must lie within $[0, 700]~\rm ns$. In addition, to exclude showers associated with charged tracks, the opening angle between the position of an EMC shower and the closest charged track at the EMC, as measured from the interaction point, must be greater than $10^\circ$. Since the signal process does not involve photons in the final state, we require the total energy of all photon candidates, $E_\gamma^{\rm total}$, to be less than $0.2~\rm GeV$ to suppress backgrounds containing neutral pions or photons. 

\begin{figure}[!ht]
    \centering
    \includegraphics[width=0.49\textwidth]{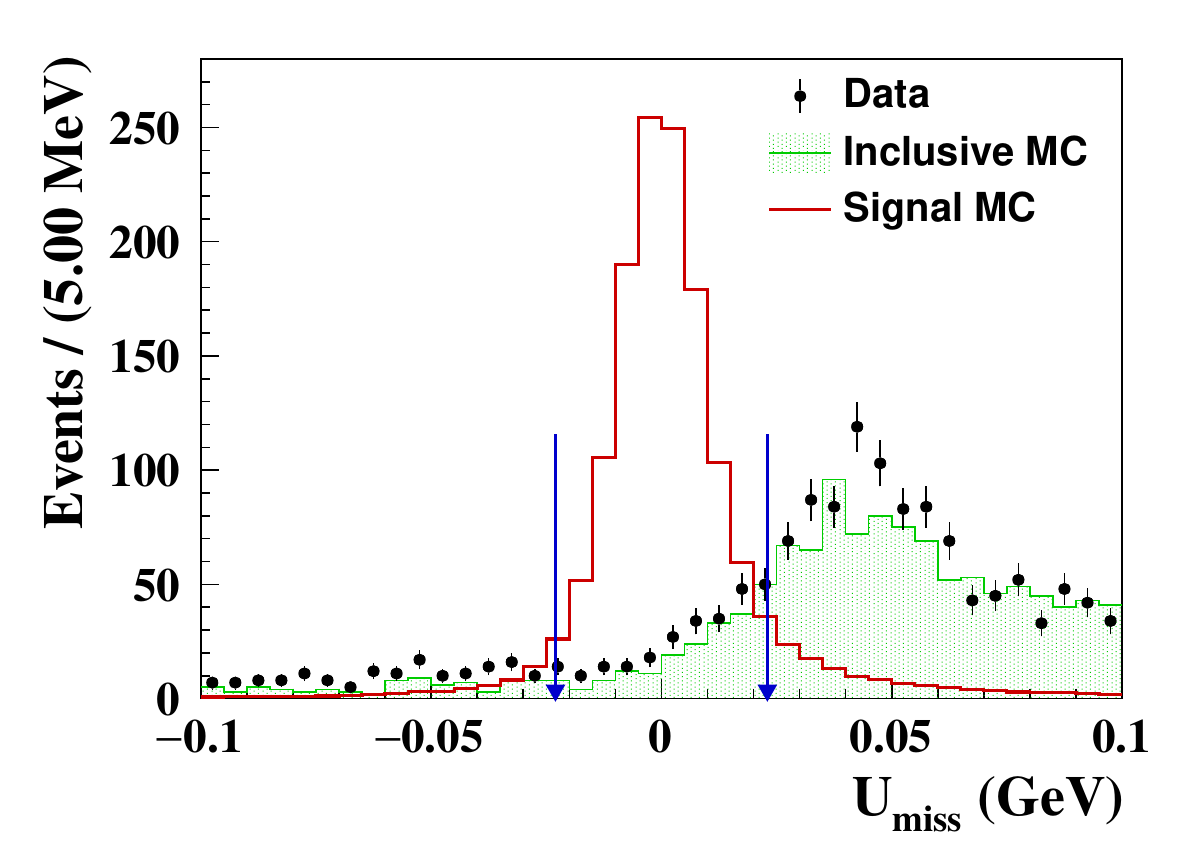}
    \caption{The $U_{\rm miss}$ distributions of the accepted candidate events. The black dots with error bars are data, the red histogram is the signal MC and the green histogram is the inclusive MC. The blue arrows mark the $|U_{\rm miss}|$ signal region.}
    \label{fg:3.1}
\end{figure}

The four-momentum of the undetected neutrino is inferred from the missing momentum and missing energy, defined as
\begin{align}
    |\vec{P}_{\rm miss}| &= \abs{\vec{0} - \sum_i\vec{P}_i},\quad i=K^-,\pi^+,K^+,e^-\\
    E_{\rm miss} &= E_{\rm cms} - \sum_i E_i,\\
    U_{\rm miss} &= E_{\rm miss} - |\vec{P}_{\rm miss}|c,
\end{align}
where $\vec{P}_{i}$ and $E_{i}$ are the three-momentum and energy of the particle $i$ in the rest frame of the initial $e^+e^-$ collision, respectively, and $E_{\rm cms}$ is the center-of-mass energy. To reduce background processes without missing particles, $|\vec{P}_{\rm miss}|$ is required to be greater than $0.05~{\rm GeV}/c$. Events with a single missing neutrino are expected to cluster around zero in the distribution of $U_{\rm miss}$. Thus, we impose the requirement $|U_{\rm miss}| < 0.023~{\rm GeV}$, as shown in Figure~\ref{fg:3.1}. 

\begin{figure}[!ht]
    \centering
    \begin{minipage}[t]{0.49\textwidth}
        \includegraphics[width=\textwidth]{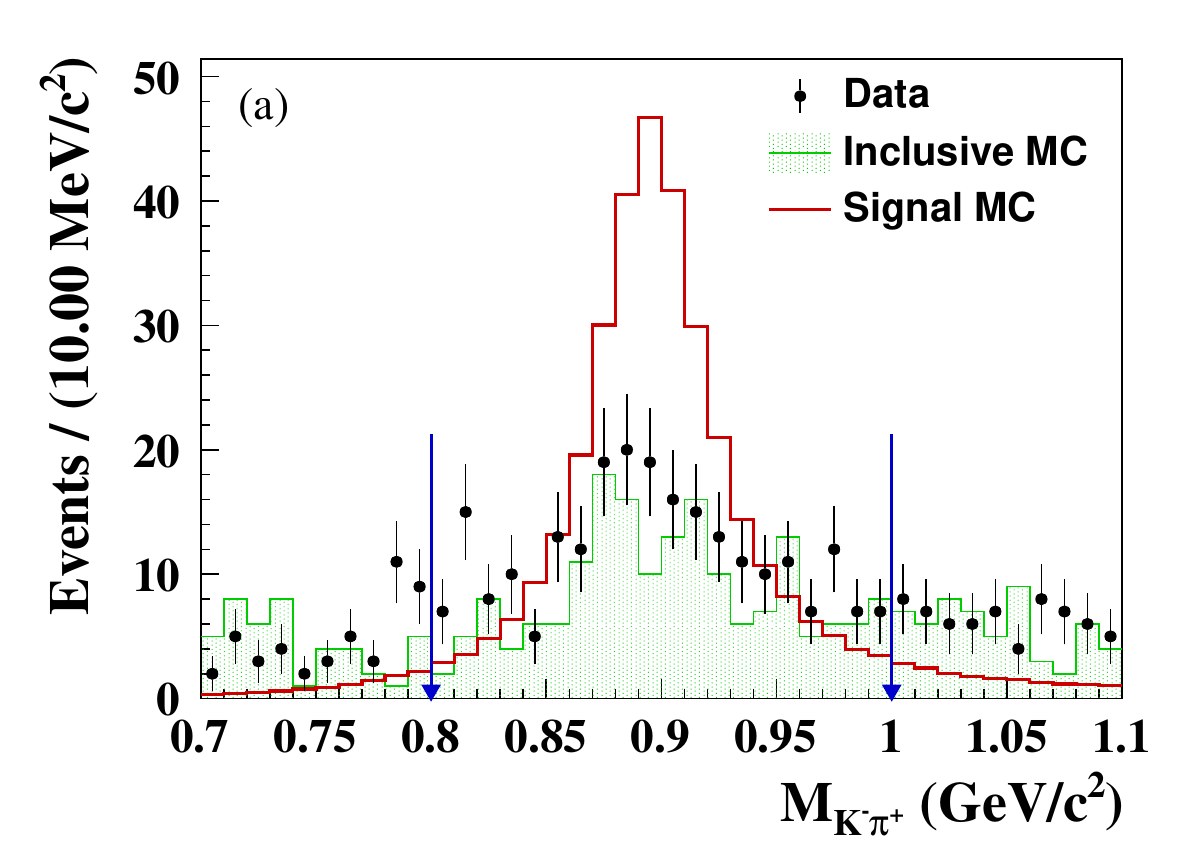}
    \end{minipage}
    \begin{minipage}[t]{0.49\textwidth}
        \includegraphics[width=\textwidth]{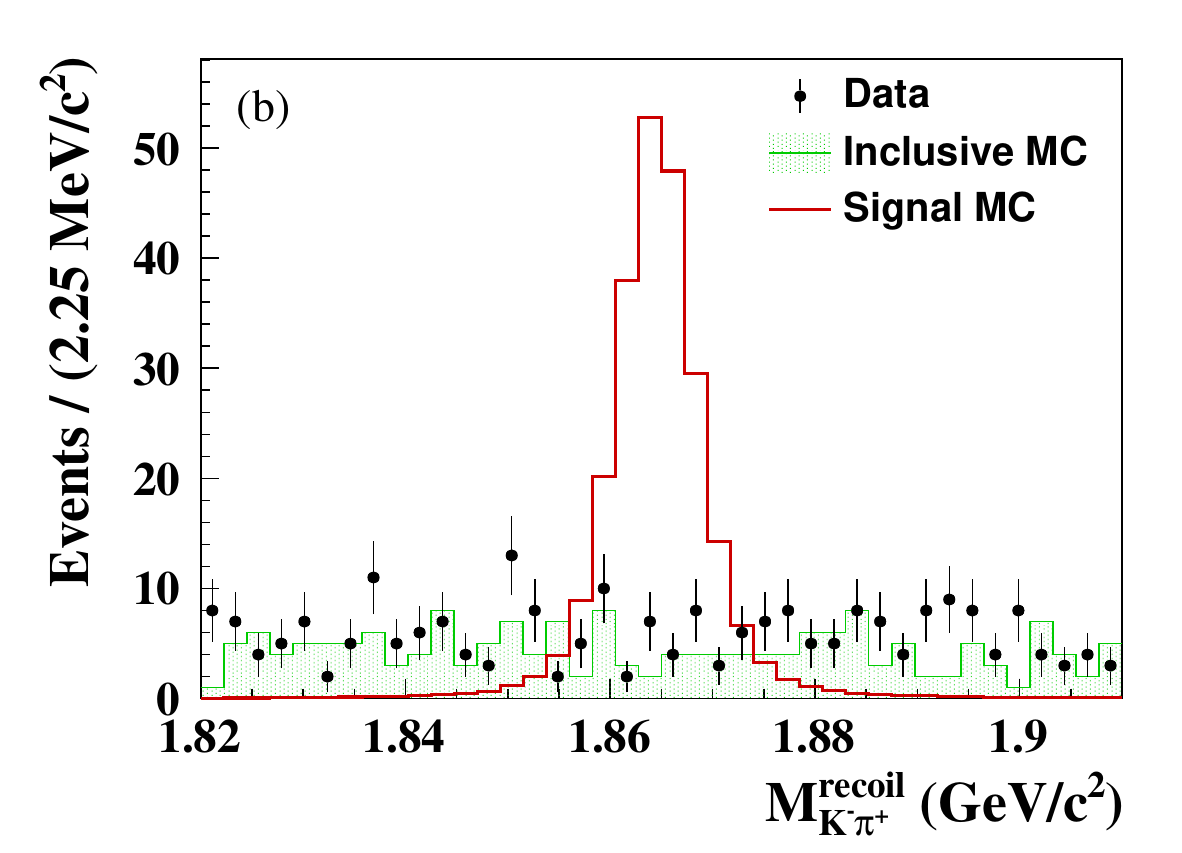}
    \end{minipage}
    \caption{The distributions of (a) $M_{K^-\pi^+}$ and (b) $M_{K^-\pi^+}^{\rm recoil}$ of the accepted candidate events. The black dots with error bars are data, the red histograms are the signal MC and the green histograms are the inclusive MC. In plot (a), the blue arrows indicate the $\bar{K}^{*0}$ signal region.}
    \label{fg:3.2}
\end{figure}

The $\bar{K}^{*0}$ meson is reconstructed via the decay $\bar{K}^{*0}\to K^-\pi^+$, with the invariant mass $M_{K^-\pi^+}$ required to be within the $\bar{K}^{*0}$ signal range defined as $(0.80,1.00)~{\rm GeV}/c^2$, corresponding to about twice the resonance width around the known $\bar{K}^{*0}$ mass. Because the BESIII detector cannot detect neutrinos, the $\bar{D}^0$ meson is identified by the recoil mass of the $K^-\pi^+$ system, $M_{K^-\pi^+}^{\rm recoil}$. The distributions of $M_{K^-\pi^+}$ and $M_{K^-\pi^+}^{\rm recoil}$ of the accepted candidate events are shown in Figure~\ref{fg:3.2}.

%%====================================================

\section{Background analysis}
\label{sec:background}

% the signal efficiency is $(24.19\pm0.07\%)$ and only 176 events remain in the inclusive MC.

% As shown in Figure~\ref{fg:47}(b), no peaking background is observed in the inclusive MC within the signal fit region $M_{K^-\pi^+}^{\rm recoil}\in(1.82,1.91)~{\rm GeV}/c^2$, simplifying signal yield fits.

After applying all the event selection criteria, the dominant background arises from the events with final states $\pi^0\pi^-\pi^+K^-K^+$ and $\gamma\pi^-\pi^+K^-K^+$. In these events, a charged pion is misidentified as an electron, and the missing momentum is attributed to a photon or a neutral pion that either escapes detection or is improperly reconstructed.

These background events pass the selection criteria through two primary mechanisms.
First, the photon may escape detection entirely by falling outside the EMC acceptance or by entering regions with reduced detection efficiency, such as the end-caps or the gaps between the barrel and end-caps. Second, for photons within the EMC barrel acceptance, their EMC showers may spatially overlap with the track projection of the misidentified pion. This additional deposited energy increases the measured $E/p$ ratio, causing the pion to mimic an electron signature.

\begin{figure}[!ht]
    \centering
    \begin{minipage}[t]{0.49\textwidth}
        \includegraphics[width=\textwidth]{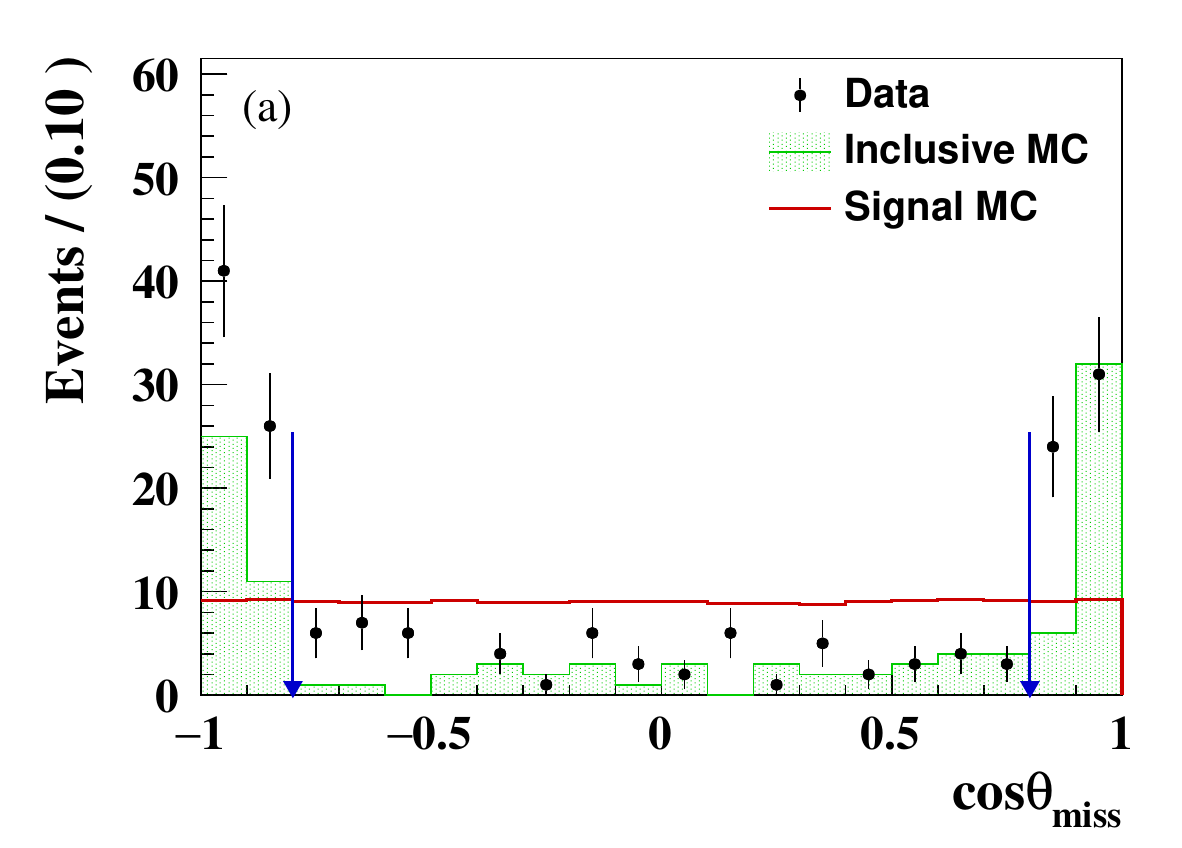}
    \end{minipage}
    \begin{minipage}[t]{0.49\textwidth}
        \includegraphics[width=\textwidth]{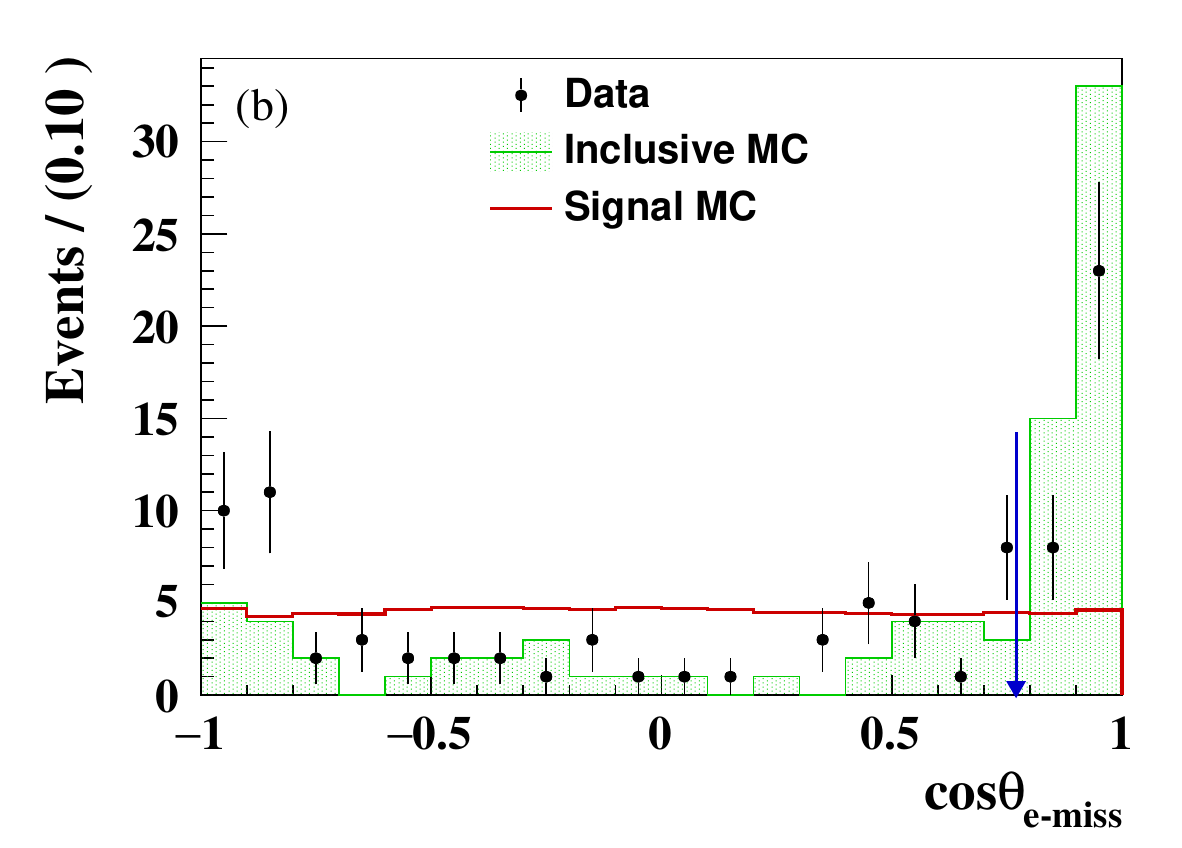}
    \end{minipage}
    \caption{The distributions of (a) $\cos\theta_{\rm miss}$ and (b) $\cos\theta_{e\text{-miss}}$ of the accepted candidate events. The black dots with error bars are data, the red histograms are the signal MC and the green histograms are the inclusive MC. The blue arrows indicate the signal region.}
    \label{fg:4.1}
\end{figure}

To effectively suppress these specific background topologies, two additional requirements are introduced. To veto events with escaped photons, the missing momentum is required to point within the reliable acceptance of the EMC barrel, namely $|\cos\theta_{\rm miss}| < 0.80$. Furthermore, to reject events featuring track-photon overlaps, the opening angle between the momentum of the electron candidate and the missing momentum is required to be sufficiently large, specifically $\cos\theta_{e\text{-miss}} < 0.77$. The distributions of these two variables are illustrated in Figure~\ref{fg:4.1}. With the inclusion of these two criteria, the final signal efficiency is determined to be $(17.31 \pm 0.06)\%$, and the background level is significantly reduced, leaving only 34 events in the inclusive MC sample.

% evades the extra shower energy requirement.

%%====================================================

\section{Result}
\label{sec:result}

To extract the signal yield, an unbinned extended maximum likelihood fit is performed on the distribution of $M_{K^-\pi^+}^{\rm recoil}$. Defining $x=M_{K^-\pi^+}^{\rm recoil}$, the complete likelihood function is given by
\begin{gather}
\mathcal{L}\qty(N_{\rm sig}, N_{\rm bkg}) = \mathcal{P}(N_{\rm obs}|N_{\rm sig}+N_{\rm bkg})\cdot\prod_i\mathcal{F}_{\rm total}\qty(x_i|N_{\rm sig}, N_{\rm bkg}), \\
\mathcal{F}_{\rm total}\qty(x|N_{\rm sig},N_{\rm bkg}) = \dfrac{N_{\rm sig}}{N_{\rm sig}+N_{\rm bkg}}{\rm PDF}_{\rm sig}\qty(x) + \dfrac{N_{\rm bkg}}{N_{\rm sig}+N_{\rm bkg}}{\rm Poly}\qty(x|c_0).
\end{gather}
In the expressions above, $\qty{x_i}$ denotes the set of $M_{K^-\pi^+}^{\rm recoil}$ values of data events and $N_{\rm obs}$ is the total number of events observed in data. The signal probability density function ${\rm PDF}_{\rm sig}(x)$ is obtained from signal MC sample, while ${\rm Poly}(x|c_0)$ represents a first-order polynomial describing the background contribution, where the coefficient $c_0$ is fixed from the inclusive MC sample. Both probability density functions are normalized. 

\begin{figure}[!ht]
\centering
\includegraphics[width=0.49\textwidth]{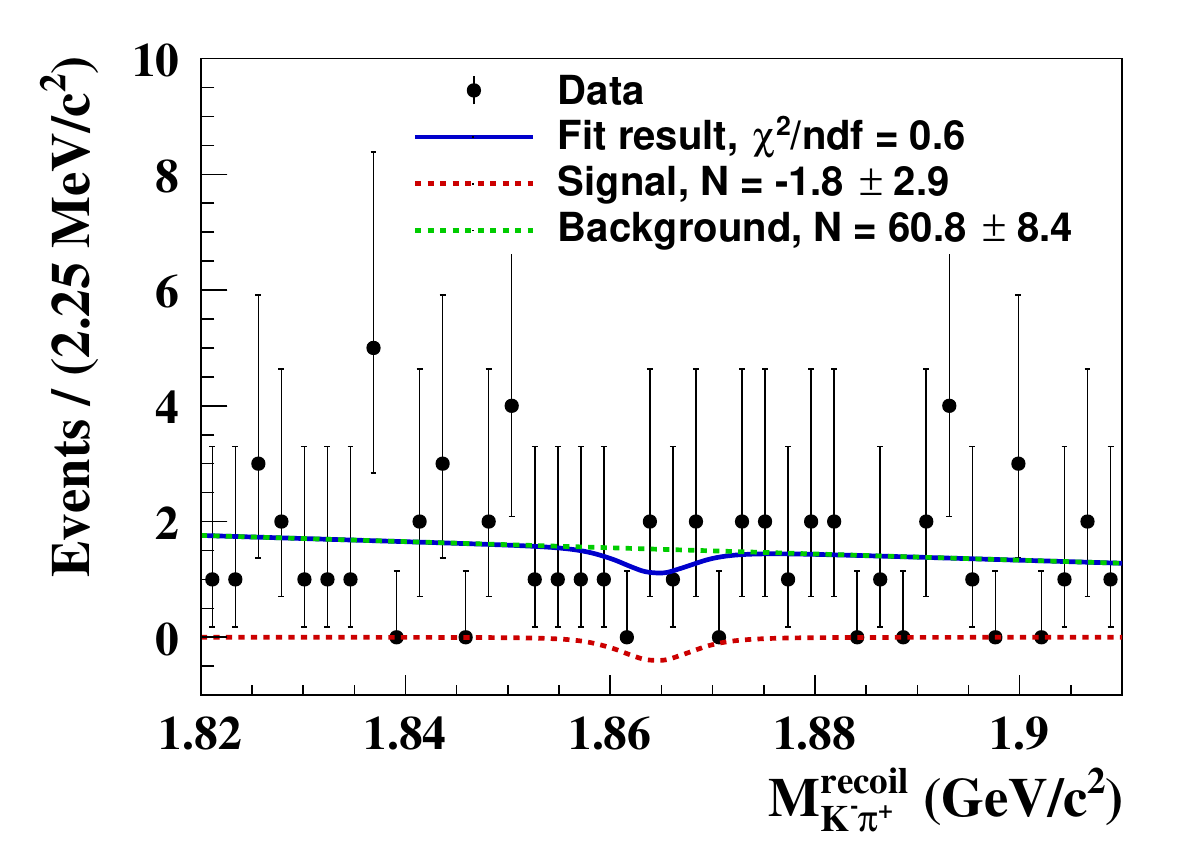}
\caption{The fit to the $M^{\rm recoil}_{K^-\pi^+}$ distribution of the accepted candidate events in data. The dots with error bars are data and the blue curve is the total fit. The red and green dashed curves represent the signal and background components, respectively.} 
\label{fg:5.1}
\end{figure}

The result of the fit to the $M_{K^-\pi^+}^{\rm recoil}$ distribution is shown in Figure~\ref{fg:5.1}. The small negative signal yield is consistent with a statistical fluctuation. When setting the upper limit (UL) on $\mathcal{B}\qty(J/\psi\to\bar{D}^0\bar{K}^{*0})$ at the 90\% CL, the non-physical region with negative signal yield will be excluded.
	
From the fit, we obtain the signal and background yields $N_{\rm sig} = -1.8\pm 2.9$ and $N_{\rm bkg} = 60.8 \pm 8.4$, indicating no significant excess above the expected background yield. Consequently, an UL on BF is set with the profile likelihood method. The BF is defined as
\begin{equation}
    \mathcal{B}(J/\psi\to\bar{D}^0\bar{K}^{*0}) = \dfrac{N_{\rm sig}}{N_{J/\psi} \, \varepsilon \, \mathcal{B}_{\rm inter}},
\end{equation}
where $N_{\rm sig},N_{J/\psi}$, and $\varepsilon$ denote the number of signal events, the total number of $J/\psi$ events in data, and the signal efficiency, respectively. The quantity $\mathcal{B}_{\rm inter}$ is defined as
    \begin{equation}\label{eq:18}
\mathcal{B}_{\rm inter} = \mathcal{B}(\bar{D}^0\to K^+e^-\bar{\nu}_e) \, \mathcal{B}(\bar{K}^{*0}\to K^-\pi^+).
    \end{equation}

\begin{figure}[!ht]
    \centering
    \includegraphics[width=0.49\textwidth]{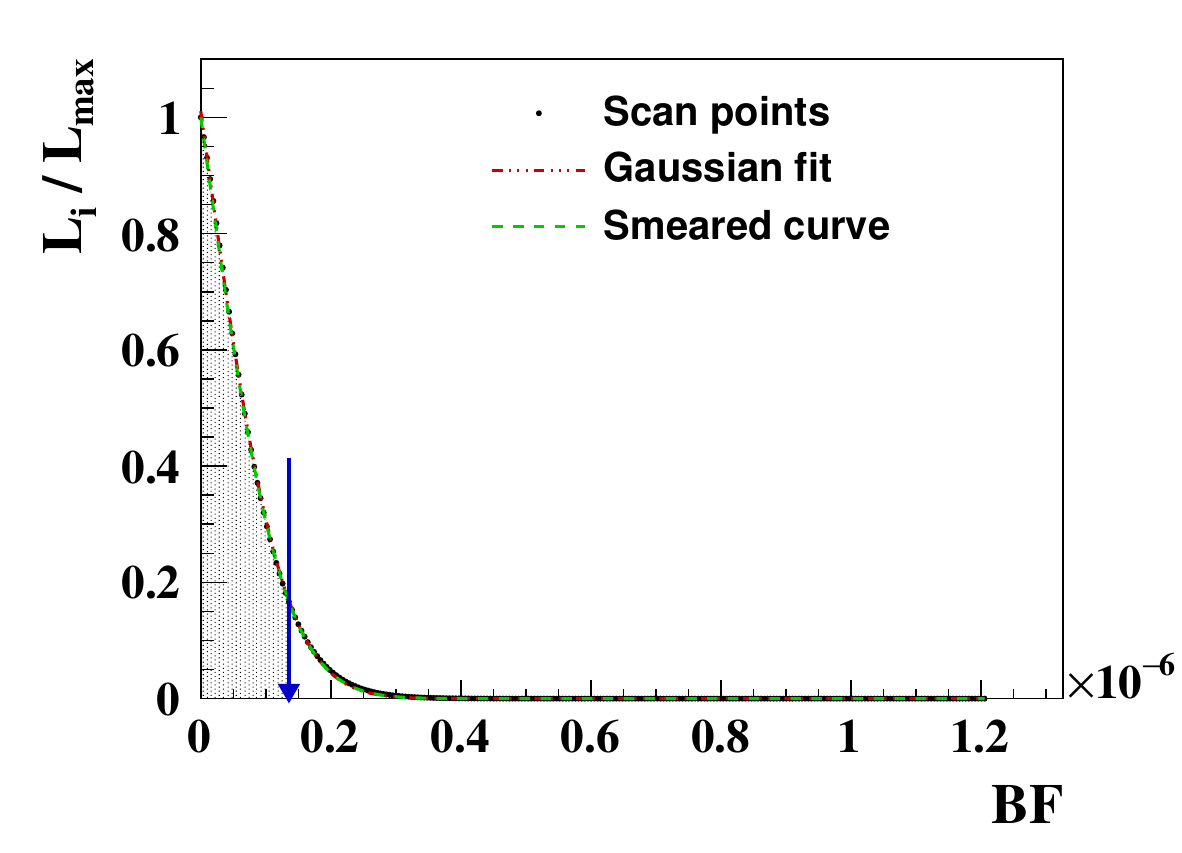}
    \caption{The likelihood ratio distribution with respect to $\mathcal{B}(J/\psi\to \bar{D}^0\bar{K}^{*0})$.  The black dots are scanned points, the red dash-dotted line line is the Gaussian fit to these points and the green dashed line is the result after convolving with an additional Gaussian function accounting for systematic uncertainties. The blue arrow indicates the UL on the BF at the 90\% CL.}
    \label{fg:5.2}
\end{figure}

For each fixed signal yield, a BF value is calculated and a likelihood value is obtained by fitting the data. A likelihood distribution is obtained by scanning the number of signal events from 0 to 50, resulting in a set of likelihood values $\mathcal{L}_i$ with a maximum value $\mathcal{L}_{\rm max}$. The relative likelihood is defined as
\begin{equation}
    \dfrac{\mathcal{L}_i}{\mathcal{L}_{\rm max}}.
\end{equation}
The relative likelihood distribution is fitted with a Gaussian function to determine the mean value $\hat{\mathcal{B}}$ and a standard deviation $\sigma_{\mathcal{B}}$ of the BF,
\begin{equation}
\mathcal{L}(\mathcal{B})_{\rm fit} \, \propto \, \exp\qty[-\dfrac{(\mathcal{B}-\hat{\mathcal{B}})^2}{2\sigma_{\mathcal{B}}^2}].
\end{equation}
Following the method of Reference~\cite{bib:7-1} for incorporating the systematic uncertainties into the UL on the BF, the likelihood function is modified to the smeared likelihood function,
\begin{equation}
    \mathcal{L}(\mathcal{B})_{\rm smear} \, \propto \, \int_0^1\exp\qty[-\dfrac{(\varepsilon\mathcal{B}/\hat{\varepsilon}-\hat{\mathcal{B}})^2}{2\sigma_{\mathcal{B}}^2}] \, \exp\qty[-\dfrac{(\varepsilon-\hat{\varepsilon})^2}{2\sigma_{\varepsilon}^2}] \, \dd\varepsilon,
\end{equation}
where $\hat{\varepsilon}$ denotes the nominal efficiency and $\sigma_{\varepsilon} = \Delta_{\rm syst} \, \varepsilon$. The relative systematic uncertainty $\Delta_{\rm syst}$ will be discussed in Section~\ref{sec:error}. The normalized likelihood versus BF is shown in Figure~\ref{fg:5.2}. By integrating the smeared likelihood curve up to 90\% of the physical region, $\mathcal{B}>0$, the UL on $\mathcal{B}(J/\psi\to\bar{D}^0\bar{K}^{*0})$ is determined to be $1.4\times10^{-7}$ at the 90\% CL, which corresponds to $N_{\rm sig}=5.7$.

%%====================================================

\section{Systematic uncertainties}
\label{sec:error}

The systematic uncertainties in the measurement of the BF of $J/\psi\to\bar{D}^0\bar{K}^{*0}$ decay mainly come from tracking, PID, the signal MC decay model, the BFs of intermediate states, the total number of $J/\psi$ events, the statistical uncertainty of signal efficiency, and the event selection criteria. The systematic uncertainties from different sources are discussed below and summarized in Table~\ref{tb:2}. The total systematic uncertainty is obtained as the sum in quadrature of all individual contributions, assuming that the systematic uncertainties are independent.

\begin{table}[!ht]
    \centering
    \caption{All systematic uncertainties in the measurement of the BF of the $J/\psi\to\bar{D}^0\bar{K}^{*0}$ decay.}
    \label{tb:2}
    \begin{tabular}{cl}
    \toprule
    Source                              & \multicolumn{1}{c}{Uncertainty/\%}\\
    \midrule
    Tracking                            & \qquad 4.0~\cite{bib:6-1,bib:6-2.2}	\\
    PID                                 & \qquad 4.0~\cite{bib:6-1,bib:6-2.2} 	\\		
    Decay model			 	            & \qquad 0.5                \\
    $\mathcal{B}_{\rm inter}$           & \qquad 0.8~\cite{pdg}     \\
    Total number of $J/\psi$ events   & \qquad 0.5~\cite{bib:3-1}	\\
    Signal MC statistics                 & \qquad 0.4                \\
    $E/p$ requirement					& \qquad 0.6		        \\
    $E_\gamma^{\rm total}$ requirement  & \qquad 0.6	            \\
    $P_{\rm miss}$ requirement 			& \qquad 0.2                \\
    $U_{\rm miss}$ requirement 	        & \qquad 1.7                \\
    $M_{K^-\pi^+}$ requirement          & \qquad 0.9	            \\
    $\cos\theta_{\rm miss}$ requirement & \qquad 0.2                \\
    $\cos\theta_{e\text{-miss}}$ requirement & \qquad 0.3           \\
    \midrule
    Total  								& \qquad{6.1}		        \\
    \bottomrule
    \end{tabular}
\end{table}

\begin{itemize}
    \item \emph{Tracking and PID.}
    The tracking and PID efficiencies for kaons and pions are studied using the control samples of the doubly tagged hadronic events of $\psi(3770)\to D^0\bar{D}^0 \, (D^+D^-)$~\cite{bib:6-1}. Specifically, the hadronic decays of $D^0\to K^-\pi^+$, $K^-\pi^+\pi^0$, $K^-\pi^-\pi^+\pi^+$ versus $\bar{D}^0\to K^+\pi^-$, $K^+\pi^-\pi^0$, $K^+\pi^+\pi^-\pi^-$, as well as $D^+\to K^-\pi^+\pi^+$ versus $D^-\to K^+\pi^-\pi^-$, are used. The electron tracking and PID efficiencies are studied using the control samples $e^+e^-\to e^+e^-(\gamma)$ and $J/\psi\to e^+e^-(\gamma)$~\cite{bib:6-2.2}. The systematic uncertainty of tracking or PID is assigned to be 1.0\% for each charged track.
    
    \item \emph{Decay model.} 
    To estimate the systematic uncertainty due to the decay model of signal simulation, an alternative phase-space model is used for comparison. The relative difference of 0.5\% between the efficiencies obtained from the nominal and phase-space models is assigned as the systematic uncertainty.

    \item \emph{BFs of $\bar{D}^0$ and $\bar{K}^{*0}$}. The uncertainty of $\mathcal{B}_{\rm inter}$ is 0.8\%~\cite{pdg}.
    
    \item \emph{Total number of ${J/\psi}$ events.}
    The total number of $J/\psi$ events, $(10087 \pm 44) \times 10^6$, is taken from Reference~\cite{bib:3-1}; its uncertainty is 0.5\%. 
    
    \item \emph{Signal MC statistics.}
    The uncertainty of the signal efficiency is estimated from the standard deviation of the Poisson-distributed selected signal yield. The resulting uncertainty is 0.4\%.

    \item \emph{$E/p$, $E_\gamma^{\rm total}$, $P_{\rm miss}$, $U_{\rm miss}$, $\cos\theta_{\rm miss}$ and $\cos\theta_{e\text{-}{\rm miss}}$ requirements.} 	
    The systematic uncertainties of the $E/p$, $E_\gamma^{\rm total}$,$P_{\rm miss}$, $U_{\rm miss}$, $\cos\theta_{\rm miss}$ and $\cos\theta_{e\text{-miss}}$ requirements are estimated to be 0.6\%, 0.6\%, 0.2\%, 1.7\%, 0.2\% and 0.3\% respectively, by using a control sample of $\psi(3770)\to\bar{D}^0D^0$, $D^0\to K^-\pi^+$, $\bar{D}^0\to K^+e^-\bar{\nu}_e$. The relative difference in efficiency between the data and MC samples is regarded as the systematic uncertainty.

    \item \emph{$M_{K^-\pi^+}$ requirements.} 
    The systematic uncertainty of the $M_{K^-\pi^+}$ requirement is evaluated as 0.9\% using the data-MC difference from a control sample of $J/\psi\to K^0\bar{K}^{*0}$, $K^0\to K_S^0\to \pi^+\pi^-$, $\bar{K}^{*0}\to K^-\pi^+$. 
\end{itemize}

%%====================================================

\section{Summary}
\label{sec:sum}

The rare weak decay $J/\psi\to\bar{D}^0\bar{K}^{*0}$ is searched for in a sample of $(10087\pm44)\times10^6$ $J/\psi$ events collected with the BESIII detector. No significant signal is observed above the expected background. The UL of its decay BF is determined to be $1.4\times10^{-7}$ at the 90\% CL, including systematic uncertainties. This UL is an order of magnitude lower than the previous best result of $2.5\times10^{-6}$~\cite{bib:1-12}. 
The new limit is still above theoretical predictions for the SM BFs of the rare $J/\psi$ decays containing a $D^0$ meson, which are of the order $10^{-9}$ or smaller.  

%%====================================================

% \appendix
% \section{Some title}
% Please always give a title also for appendices.

%%====================================================

\acknowledgments

The BESIII Collaboration thanks the staff of BEPCII (https://cstr.cn/31109.02.BEPC) and the IHEP computing center for their strong support. This work is supported in part by National Key R\&D Program of China under Contracts Nos. 2023YFA1606000, 2023YFA1606704; National Natural Science Foundation of China (NSFC) under Contracts Nos. 11635010, 11935015, 11935016, 11935018, 12025502, 12035009, 12035013, 12061131003, 12192260, 12192261, 12192262, 12192263, 12192264, 12192265, 12221005, 12225509, 12235017, 12361141819; the Chinese Academy of Sciences (CAS) Large-Scale Scientific Facility Program; CAS under Contract No. YSBR-101; 100 Talents Program of CAS; The Institute of Nuclear and Particle Physics (INPAC) and Shanghai Key Laboratory for Particle Physics and Cosmology; ERC under Contract No. 758462; German Research Foundation DFG under Contract No. FOR5327; Istituto Nazionale di Fisica Nucleare, Italy; Knut and Alice Wallenberg Foundation under Contracts Nos. 2021.0174, 2021.0299; Ministry of Development of Turkey under Contract No. DPT2006K-120470; National Research Foundation of Korea under Contract No. NRF-2022R1A2C1092335; National Science and Technology fund of Mongolia; Polish National Science Centre under Contract No. 2024/53/B/ST2/00975; STFC (United Kingdom); Swedish Research Council under Contract No. 2019.04595; U. S. Department of Energy under Contract No. DE-FG02-05ER41374

    \bibliographystyle{JHEP}
    \bibliography{Pubcom_biblio.bib}

\section*{The BESIII Collaboration}\label{author}
\noindent
M.~Ablikim$^{1}$\BESIIIorcid{0000-0002-3935-619X},
M.~N.~Achasov$^{4,b}$\BESIIIorcid{0000-0002-9400-8622},
P.~Adlarson$^{77}$\BESIIIorcid{0000-0001-6280-3851},
X.~C.~Ai$^{82}$\BESIIIorcid{0000-0003-3856-2415},
R.~Aliberti$^{36}$\BESIIIorcid{0000-0003-3500-4012},
A.~Amoroso$^{76A,76C}$\BESIIIorcid{0000-0002-3095-8610},
Q.~An$^{73,59,\dagger}$,
Y.~Bai$^{58}$\BESIIIorcid{0000-0001-6593-5665},
O.~Bakina$^{37}$\BESIIIorcid{0009-0005-0719-7461},
Y.~Ban$^{47,g}$\BESIIIorcid{0000-0002-1912-0374}
% ...
\\
% \vspace{0.2cm}
% (BESIII Collaboration)\\
\vspace{0.2cm} {\it\\
$^{1}$ Institute of High Energy Physics, Beijing 100049, People's Republic of China\\
% ...
$^{4}$ Budker Institute of Nuclear Physics SB RAS (BINP), Novosibirsk 630090, Russia\\
% ...
$^{36}$ Johannes Gutenberg University of Mainz, Johann-Joachim-Becher-Weg 45, D-55099 Mainz, Germany\\
$^{37}$ Joint Institute for Nuclear Research, 141980 Dubna, Moscow region, Russia\\
% ...
$^{47}$ Peking University, Beijing 100871, People's Republic of China\\
% ...
$^{58}$ Southeast University, Nanjing 211100, People's Republic of China\\
$^{59}$ State Key Laboratory of Particle Detection and Electronics, Beijing 100049, Hefei 230026, People's Republic of China\\
% ...
$^{73}$ University of Science and Technology of China, Hefei 230026, People's Republic of China\\
% ...
$^{76}$ University of Turin and INFN, (A)University of Turin, I-10125, Turin, Italy; (B)University of Eastern Piedmont, I-15121, Alessandria, Italy; (C)INFN, I-10125, Turin, Italy\\
$^{77}$ Uppsala University, Box 516, SE-75120 Uppsala, Sweden\\
% ...
$^{82}$ Zhengzhou University, Zhengzhou 450001, People's Republic of China\\

\noindent
$^{\dagger}$ Deceased\\
% ...
$^{b}$ Also at the Novosibirsk State University, Novosibirsk, 630090, Russia\\
% ...
$^{g}$ Also at State Key Laboratory of Nuclear Physics and Technology, Peking University, Beijing 100871, People's Republic of China\\
% ...
}

\end{document}